\documentclass[aps,prb,twocolumn,showpacs,floatfix,groupedaddress,superscriptaddress]{revtex4-2}
\pdfoutput=1
\usepackage[linkcolor=blue,colorlinks=true,breaklinks=true,citecolor=blue,urlcolor=blue]{hyperref}
\usepackage{graphicx,graphics} 
\usepackage{physics,amsmath,amssymb,amsbsy}
\usepackage{braket}
\usepackage{bm}
\usepackage{dcolumn}
\usepackage{dsfont}
\usepackage[mathscr]{euscript}
\usepackage{xcolor}
\usepackage[titletoc,title]{appendix}
\usepackage[dotinlabels]{titletoc}
\usepackage[caption=false,labelformat=simple]{subfig}

\begin{document}
\title{Layer-Dependent Orbital Magnetization in Graphene-Haldane Heterostructures}
\author{Sovan Ghosh}
\email{sovanghosh2014@gmail.com}
\affiliation{Department of Physical Sciences, Indian Institute of Science Education and Research Kolkata\\ Mohanpur-741246, West Bengal, India}
\author{Bheema Lingam Chittari}
\email{bheemalingam@iiserkol.ac.in}
\affiliation{Department of Physical Sciences, Indian Institute of Science Education and Research Kolkata\\ Mohanpur-741246, West Bengal, India}
\begin{abstract}
Rhombohedral multilayer graphene (RMG) proximity-coupled to a Haldane substrate provides a platform to investigate the interplay between band topology, layer number, and electric-field control of orbital magnetism. Using a tight-binding model and the modern theory of orbital magnetization, we study the layer-dependent magnetic response in bilayer, trilayer, tetralayer and pentalayer graphene under Haldane proximity. While monolayer graphene develops a global topological gap with quantized magnetization slope, multilayer systems remain metallic due to protected low-energy bands associated with unperturbed sublattices. Despite the absence of a global gap, finite valley-contrasting Berry curvature produces non-trivial layer-dependent Chern numbers. We decompose the total orbital magnetization into self-rotation ($M_{\mathrm{SR}}$) and center-of-mass ($M_C$) contributions, revealing their distinct behaviors across doping and applied interlayer bias. In bilayer graphene, magnetization remains negative and monotonic. Remarkably, trilayer and tetralayer graphene display a bias-induced sign reversal of orbital magnetization beyond critical thresholds ($\Delta \simeq -55$ meV for 3LG, $-50$ meV for 4LG) in the hole-doped regime, a feature completely absent in the bilayer. It is further established in the case of 5LG, that the magnetization reversal is independent of the topological transition, and depends on the direction of bias and hole doping. The effect persists across both hole and electron doping, demonstrating that layer count serves as a key tuning parameter for orbital magnetism. Our findings establish topologically proximitized multilayer graphene as a versatile platform for electric-field-manipulable orbitronic and valleytronic devices.
\end{abstract}

\maketitle
\section{\label{sec:1}Introduction}
Rhombohedral multilayer graphene (RMG) has emerged as a rich platform for exploring correlated and topological phenomena due to its eightfold spin-valley-sublattice degeneracy. Lifting this degeneracy through symmetry breaking enables a variety of ordered states, with orbital magnetism standing out as a particularly striking consequence of non-trivial band geometry and Berry curvature effects~\cite{Zhou2021nat,PhysRevLett.106.156801,Xiao2007}. Unlike conventional spin magnetism, orbital magnetization arises from self-rotations and center-of-mass motions of Bloch wave packets, making it highly sensitive to the underlying band topology~\cite{Xiao2007,Geisenhof2021,Huang2023,Mainak2024,Arp2024}. Considerable experimental progress has been made in twisted and multilayer graphene systems coupled to symmetry-breaking substrates such as hexagonal boron nitride (hBN) and transition metal dichalcogenides (TMDs)~\cite{Han2023,Liu2019,Tschirhart2021,Sharpe2021,Liu2021,Bhowmik2023}. In twisted bilayer graphene aligned with hBN, and in rhombohedral trilayer graphene aligned with hBN, spontaneous breaking of time-reversal and inversion symmetries gives rise to valley-contrasting Chern bands and gate-tunable anomalous Hall effects~\cite{Polshyn2020,Sharpe2021,Serlin2020}. These discoveries established that orbital magnetism in moiré systems is not merely a passive response but an actively tunable property linked to topological edge states~\cite{He2020,PhysRevLett.125.227702}. More recently, the landscape has expanded beyond moiré platforms: experiments on rhombohedral hexalayer graphene have revealed multiferroic orbital magnetism with electric-field-reversible hysteresis, characterized by a $\Delta P \cdot M$ order parameter, demonstrating that these phenomena persist and diversify in higher-order rhombohedral stacks~\cite{Deng2025arXiv}. More notably, the exotic properties and time-reversal symmetry breaking in RMG have led to the discovery of several remarkable phenomena, including the quantum anomalous Hall effect, fractional quantum anomalous Hall effect, superconductivity, and signatures of chiral superconductivity~\cite{Lu2024, Choi2025, Han2025, gm64-vxdm}. In contrast to these moir\'e or intrinsic RMG systems, our work focuses on RMG proximitized by a Haldane substrate, where orbital magnetization and magnetic reversal can be controlled purely by layer number, carrier density, and gate voltage without the need for moir\'e superlattices. These experimental advances have been accompanied by intensive theoretical efforts exploring the underlying mechanisms, including the role of quantum geometry and Coulomb interactions in stabilizing fractional phases~\cite{Dong2024, Miao2025}, the emergence of nematic and partially polarized phases~\cite{Parra-Martinez2025}, analytic models of Berry curvature and Wigner crystallization~\cite{Bernevig2025}, and proposals for Majorana chiral superconductivity~\cite{Yoon2026}.  
\par Parallel to these experimental advances, proximity engineering has emerged as a powerful alternative to moiré patterning for inducing topological order. A Haldane-like state with non-zero Chern number has recently been realized in monolayer graphene via an exchange field from a magnetic substrate, establishing a new pathway to quantum anomalous Hall phases without external magnetic fields~\cite{Domaretskiy2025ProximityScreening}. The microscopic origin of such proximity effects lies in resonant band hybridization across van der Waals interfaces. For instance, alignment between the $Q$-point conduction band of WS$_2$ and the Dirac cone of graphene mediates interlayer coupling, enhancing valley Zeeman splitting and modifying orbital $g$-factors~\cite{FariaJunior2023}. Complementary theoretical work has demonstrated that simultaneous proximity spin-orbit and exchange fields in trilayer RMG stabilize competing correlated phases, with the ground state sensitively dependent on the relative magnetization orientation of encapsulating ferromagnetic layers~\cite{Zhumagulov2024PRL}. These findings establish that proximity effects operate not merely as perturbative symmetry-breaking fields but as resonant orbital hybridization phenomena capable of engineering non-trivial band topology. Despite this progress, the layer-dependent evolution of orbital magnetism in proximitized RMG systems remains largely unexplored. While monolayer graphene under Haldane proximity develops a global topological gap with quantized magnetization response, multilayer systems retain metallic character due to protected low-energy bands originating from unperturbed sublattices. This raises fundamental questions: How does orbital magnetization evolve with increasing layer number? Can electric-field control be extended to multi-layer stacks, and what new phenomena emerge from the interplay between layer count, band topology, and applied bias?

\par In this work, we address these questions by investigating orbital magnetization in topologically proximitized few-layer RMG (bilayer, trilayer, tetralayer, and pentalayer) using a tight-binding framework and the modern theory of orbital magnetization. We uncover a distinct layer-dependent phenomenon: bias-induced sign reversal of orbital magnetization in trilayer, tetralayer and pentalayer systems. Beyond a critical displacement field, weakly hole-doped and electron-doped regimes exhibit opposite magnetization signs, a feature completely absent in bilayer graphene. Our results establish layer number as a critical control parameter for orbital magnetism and position RMG as a versatile platform for electrically switchable orbitronic and valleytronic devices.
\begin{figure*}[t]
    \centering
    \includegraphics[width=\textwidth]{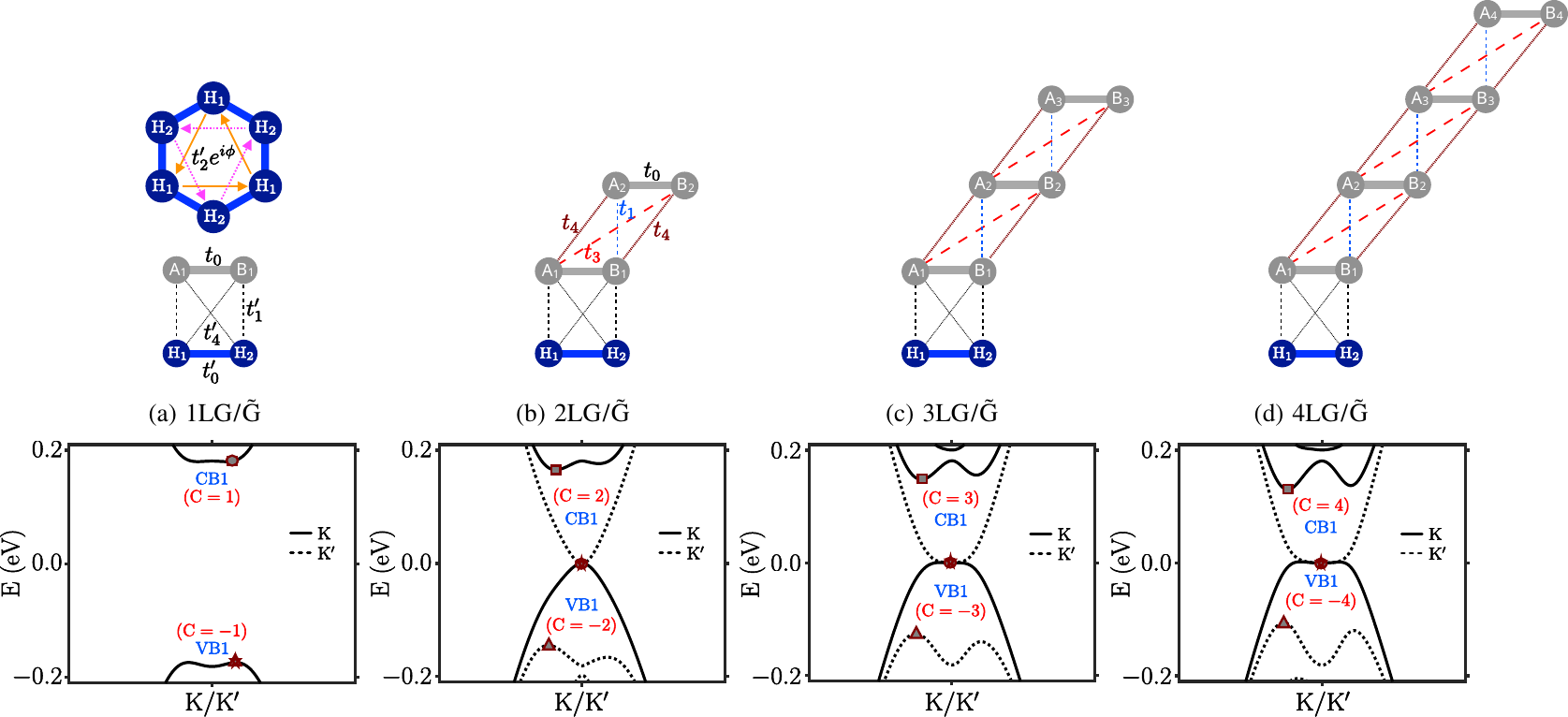}
    \caption{(a)-(d) Low-energy bands along with schematic diagram of N-layer graphene (NLG) on a Haldane layer ($\tilde{G}$) at zero gate voltage in graphene layers. The blue-filled circles $H_1$ and $H_2$ in the schematic diagram represent the two distinct sublattices of the Haldane substrate layer. Intra-layer hopping within the Haldane layer is represented by $t^{\prime}_0$, and inter-layer hopping between the Haldane layer and the proximity layer of NLG is represented by $t^{\prime}_1$ and $t^{\prime}_4$. The next-nearest-neighbor (NNN) complex hopping process in the Haldane layer is represented by $t_2^{\prime}e^{i\phi}$. The gray-filled circles labeled $(A_i, B_i)$ represent the two sublattices of each graphene layer in NLG. Solid and dashed black lines represent the K and $K^{\prime}$ valley-resolved low-energy bands, respectively. The triangle, pentagon, circle, and square markers indicate the band edges of VB1 at $K^{\prime}$, VB1 at $K$, CB1 at $K^{\prime}$, and CB1 at $K$, respectively. The Chern number of the individual bands is also equal to the number of graphene layers in the system.}
    \label{multi_bands}
\end{figure*}
\section{{\label{sec:method}}Low Energy Spectrum in Haldane-Proximitized Multilayer Graphene}

The single-particle tight-binding Hamiltonian in momentum space for rhombohedral N-layer graphene (NLG) in proximity to a Haldane layer ($\mathrm{NLG}/\tilde{G}$) is given by~\cite{Ghosh_2025}:
\begin{equation}\label{Hamil}
    H(\vb{k}) = H^{\mathrm{NLG}} + H^h + T
\end{equation}

Here, $H^{\mathrm{NLG}}$, $H^h$, and $T$ denote the Hamiltonians for the N-layer graphene, the Haldane substrate, and their interlayer coupling, respectively. The explicit form of $H^{\mathrm{NLG}}$, written in the ($A_1,B_1,A_2,B_2,...,A_N,B_N$) sublattice basis, is provided in Appendix \ref{app.NLG}. The substrate is described by the Haldane Hamiltonian~\cite{sovan2025,Haldane}, $H^h(\vb{k})=\vb{d^h(k)\cdot\sigma} + \vb{\epsilon^h} \cdot \mathbb{I}$, where $\sigma$ are Pauli matrices, $d_x^h=t^{\prime}_0\sum_{i=1}^3 cos(\vb{k\cdot e_i})$, $d_y^h=-t^{\prime}_0\sum_{i=1}^3 sin(\vb{k\cdot e_i})$, $d_z^h=M-2t_2^{\prime} sin(\phi)\sum_{i=1}^3 sin(\vb{k\cdot b_i})$, and $\epsilon^h=2t_2^{\prime} cos(\phi)\sum_{i=1}^3 cos(\vb{k\cdot b_i})$. $\vb{b_1=e_2-e_3}$, $\vb{b_2=e_3-e_1}$, and $\vb{b_3=e_1-e_2}$ are the next-nearest neighbour connecting vectors, where $\vb{e_1}$, $\vb{e_2}$, and $\vb{e_3}$ are connecting the nearest neighbour. Here, $M$ is the Semenoff mass (on-site potential), $t_{2}^{\prime}$ is the next-nearest-neighbor (NNN) hopping strength, $\phi$ is the NNN hopping phase, and $t^{\prime}_0$ is the nearest-neighbor (NN) hopping strength in Haldane substrate. The Haldane layer is positioned directly beneath the graphene and couples only to its bottom layer. In this model, we assume perfect lattice matching. The coupling matrix $T$ between the Haldane substrate and the graphene's bottom layer is
\begin{equation}
T = 
\begin{bmatrix}
t_1^{\prime} & t_4^{\prime} f(\vb{k}) \\
t_4^{\prime} f^{\dagger}(\vb{k}) & t_1^{\prime}
\end{bmatrix}.
\end{equation}

Here, $t_1^{\prime}$ and $t_4^{\prime}$ are the vertical and skew interlayer hopping strengths, respectively. The structure factor $f(\vb{k})$ is defined as $f(\vb{k}) = \sum_{j=1}^3 e^{i\vb{k} \cdot \vb{e}_j}$, where $\vb{e}_j$ are vectors connecting nearest neighbors. To preserve the tunable topological bands, we assume that the interlayer coupling strengths between the bottom graphene layer and the Haldane substrate ($t_1^{\prime}$ and $t_4^{\prime}$) are identical to those within rhombohedrally stacked graphene layers ($t_1$ and $t_4$). While our calculations assume commensurate stacking and set interlayer couplings $t_1'$ and $t_4'$ equal to those within graphene for simplicity, the qualitative features, layer-dependent band protection and valley-contrasting Berry curvature, are expected to persist for moderate variations in these parameters. In realistic samples, lattice mismatch or twist would introduce moiré superlattices~\cite{PhysRevLett.122.016401}, potentially modifying quantitative details; however, the essential physics of bias-controlled orbital magnetization should survive as long as the interlayer coupling remains sufficiently strong to maintain proximity-induced topology. Following our previous work~\cite{Ghosh_2025}, we set the Haldane parameters as $t_2^{\prime}=0.1$ eV, phase $\phi=\pi/2$, and mass $M=0$. The low-energy band structure of the proximity-coupled graphene layers is presented in Fig.~\ref{multi_bands}. As established in the monolayer case, proximity-induced symmetry breaking, specifically an effective sublattice-dependent onsite potential, opens a well-defined local band gap at the Dirac points ($\rm K$ and $\rm K^{\prime}$) in single-layer graphene (1LG/$\tilde{G}$). This gap appears at charge neutrality due to the lifting of sublattice degeneracy. In contrast, a different electronic landscape emerges in rhombohedral (ABC) stacked multilayer graphene. In these systems, the substrate-induced potential does not affect all atoms equally. Due to the stacking geometry, the $\rm{A_1}$ sublattice in the bottom (proximity-coupled) layer is directly perturbed, while the $\rm{B_N}$ sublattice in the top layer remains largely shielded. More importantly, the unperturbed sublattices from the outer layers give rise to low-energy bands that remain nearly degenerate at the Fermi level. At a given valley ($\rm K$ or $\rm K^{\prime}$), these protected bands appear as either valence or conduction bands that pin the Fermi energy, preventing the establishment of a global insulating gap. This behavior is clearly visible in the calculated band structures for bilayer, trilayer, and tetralayer rhombohedral stacks, shown in Fig.~\ref{multi_bands}(b)-(d). As a result, proximity-coupled N-layer graphene ($\rm{N}$LG) remains metallic, in contrast to the gapped, insulating nature of proximity-coupled monolayer (1LG/$\tilde{G}$). This distinction highlights the crucial role of interlayer coupling and stacking order in modulating proximity effects. Despite the absence of a global gap, these low-energy bands become topologically non-trivial due to the Haldane substrate. At zero bias voltage, the Chern number of each low-energy band is proportional to the number of graphene layers, as indicated in Fig.~\ref{multi_bands}.
\section{\label{sec:3}Orbital Magnetization in Haldane-Proximitized Multilayer Graphene}
The orbital magnetization of the electronic bands is calculated using the modern theory of orbital polarization, which is well-suited for systems with non-trivial Berry curvature~\cite{sovan2025, PhysRevLett.106.156801}. The total orbital magnetization can be decomposed into two physically distinct contributions: one from the self-rotation of wave packets, $M_{\mathrm{SR}}$, and another from their center-of-mass motion, $M_{C}$. The total orbital magnetization is given by:
\begin{equation}
\resizebox{0.95\hsize}{!}{$
\begin{aligned}
M_z^{\mathrm{orb}} = -\frac{e}{\hbar} \int \frac{d^2k}{(2\pi)^2} \sum_n f_n(\mathbf{k})\sum_{n' \ne n} \left[ (E_n - E_{n'}) + 2(\mu - E_n) \right] \\ \times\, \mathrm{Im} \left( \frac{\langle \psi_n | \frac{\partial H}{\partial k_x} | \psi_{n'} \rangle\langle \psi_{n'} | \frac{\partial H}{\partial k_y} | \psi_n \rangle}{(E_n -E_{n'})^2}\right)
\end{aligned}
$}
\end{equation}
The two components, $M_{\mathrm{SR}}$ and $M_{C}$, can be expressed in terms of the band-resolved Berry curvature $\Omega_n^{xy}(\mathbf{k})$:
\begin{equation*}
M_{\mathrm{SR}}=\frac{e}{\hbar} \int \frac{d^2k}{(2\pi)^2}\sum_n f_n(k)\frac{(E_n-E_{n^{\prime}})}{2}\Omega_n^{xy}
\end{equation*} 
and 
\begin{equation*}
M_C=\frac{e}{\hbar} \int \frac{d^2k}{(2\pi)^2}\sum_n f_n(k)(\mu-E_n)\Omega_n^{xy},
\end{equation*}
where the Berry curvature is defined as:
\begin{equation*}
\Omega_n^{xy}=-2\sum_{n^{\prime}\neq n}{\rm Im}\left[\frac{\bra{\psi_n}\frac{\partial H}{\partial k_x}\ket{\psi_{n^{\prime}}}\bra{\psi_{n^{\prime}}}\frac{\partial H}{\partial k_y}\ket{\psi_n}}{(E_n-E_{n^{\prime}})^2}\right].
\end{equation*}
In our calculations, we restrict the states to the low-energy valence and conduction bands near charge neutrality, as they are the primary contributors to the topological response.
\subsection{Magnetization in $1\mathrm{LG}/\tilde{G}$}
In systems with broken time-reversal and inversion symmetries, spontaneous valley polarization can lead to orbital magnetic states whose strength and sign are directly controlled by band topology. For a system with a global band gap, the slope of the orbital magnetization with respect to the Fermi energy satisfies $\frac{dM_z}{dE_f} = \frac{Ce}{2\pi \hbar}$~\cite{PhysRevLett.125.227702}, directly linking magnetization to the Chern number.
\begin{figure}[t]
    \centering
    \includegraphics[width=0.8\columnwidth]{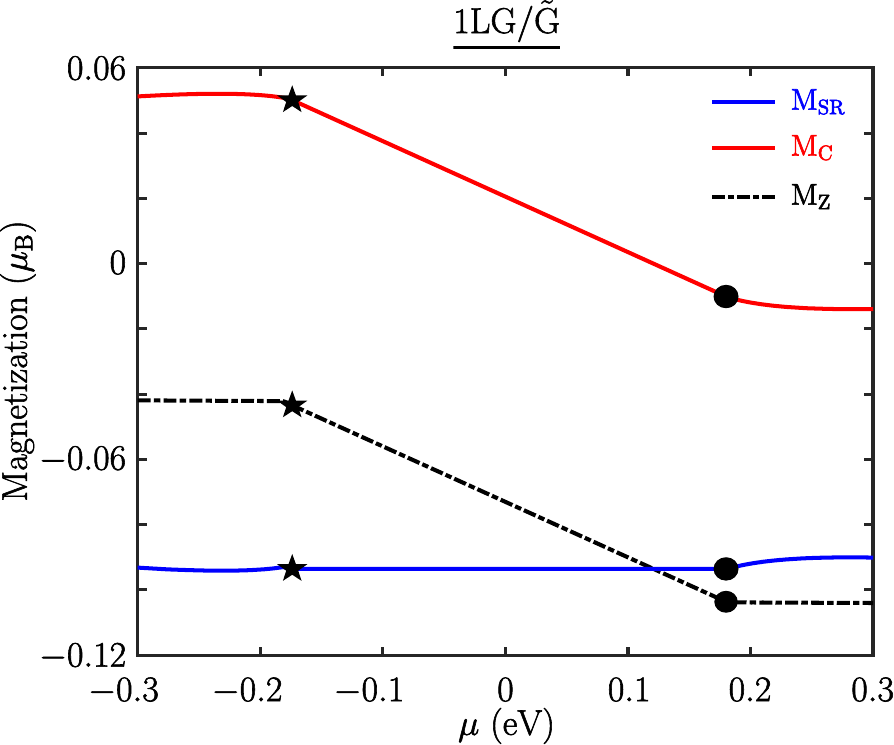}
    \caption{Orbital magnetization in monolayer graphene (1LG) on a Haldane layer ($\tilde{G}$). The total orbital magnetization (black dashed line) is decomposed into contributions from self-rotation ($M_{\mathrm{SR}}$, blue line) and center-of-mass motion ($M_C$, red line). The star and dot mark the valence band maximum and conduction band minimum, respectively.}
    \label{1LG_OrbMag}
\end{figure}
The low-energy bands [Fig.~\ref{multi_bands}(a)] are topologically non-trivial, characterized by a nonzero Chern number ($C=\pm1$) and a net orbital magnetization $M_z^{\mathrm{orb}}$ [Fig.~\ref{1LG_OrbMag}]. The two contributions behave distinctly: $M_{\mathrm{SR}}$ remains nearly constant at approximately $-93.5 \times 10^{-3}\,\mu_B$ across the band gap, while $M_C$ varies linearly within the gap, consistent with the topological formula. This clear separation in the gapped monolayer provides a foundation for understanding the more complex magnetization response in multilayer structures.
\begin{figure}[t]
    \centering
    \includegraphics[width=0.8\columnwidth]{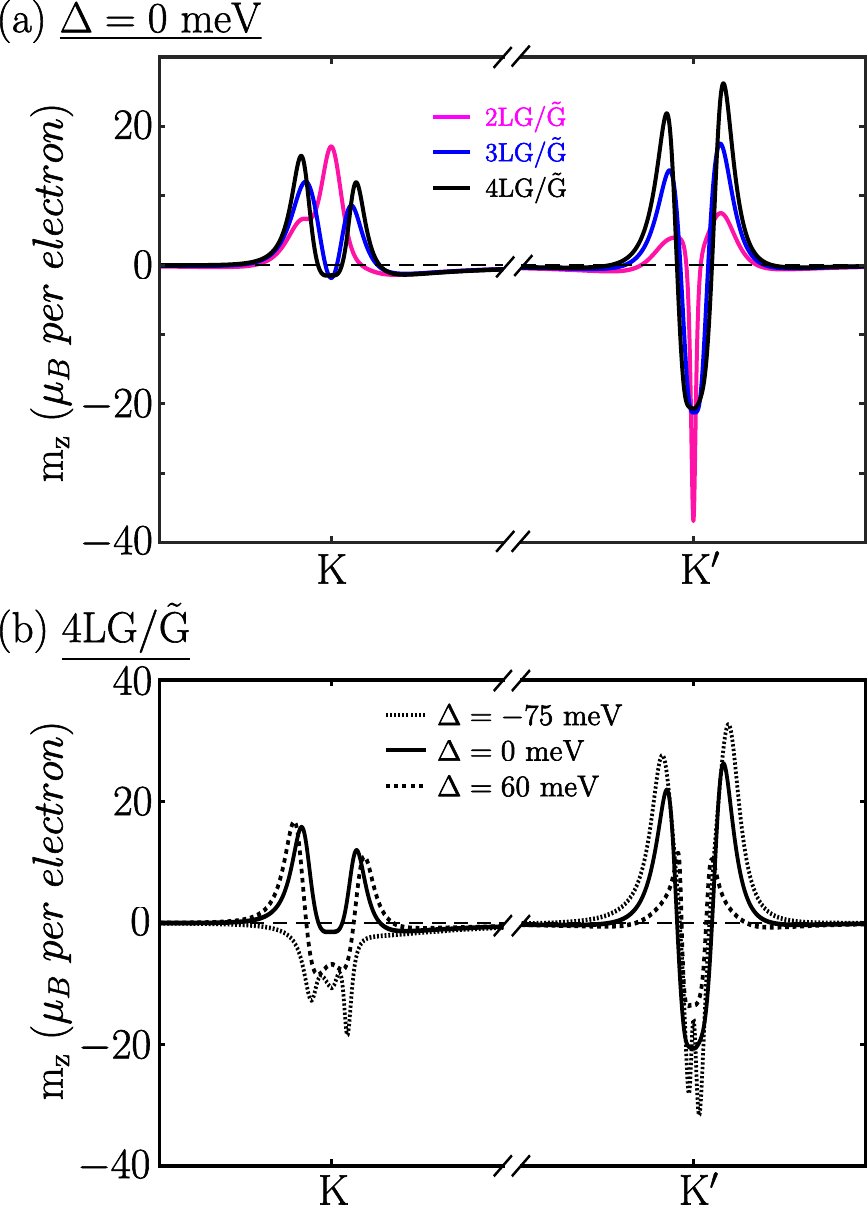}
    \caption{Orbital magnetic moment distribution of the valence band at zero bias voltage at valleys $K$ and $K^{\prime}$ with increasing number of graphene layers (a) and with different interlayer potential bias at a $\rm{4LG/\tilde{G}}$ (b).}
    \label{mz_line}
\end{figure}
\subsection{Layer-Dependent Magnetization in Multilayer Graphene}
\begin{figure}[t]
    \centering
    \includegraphics[width=\columnwidth]{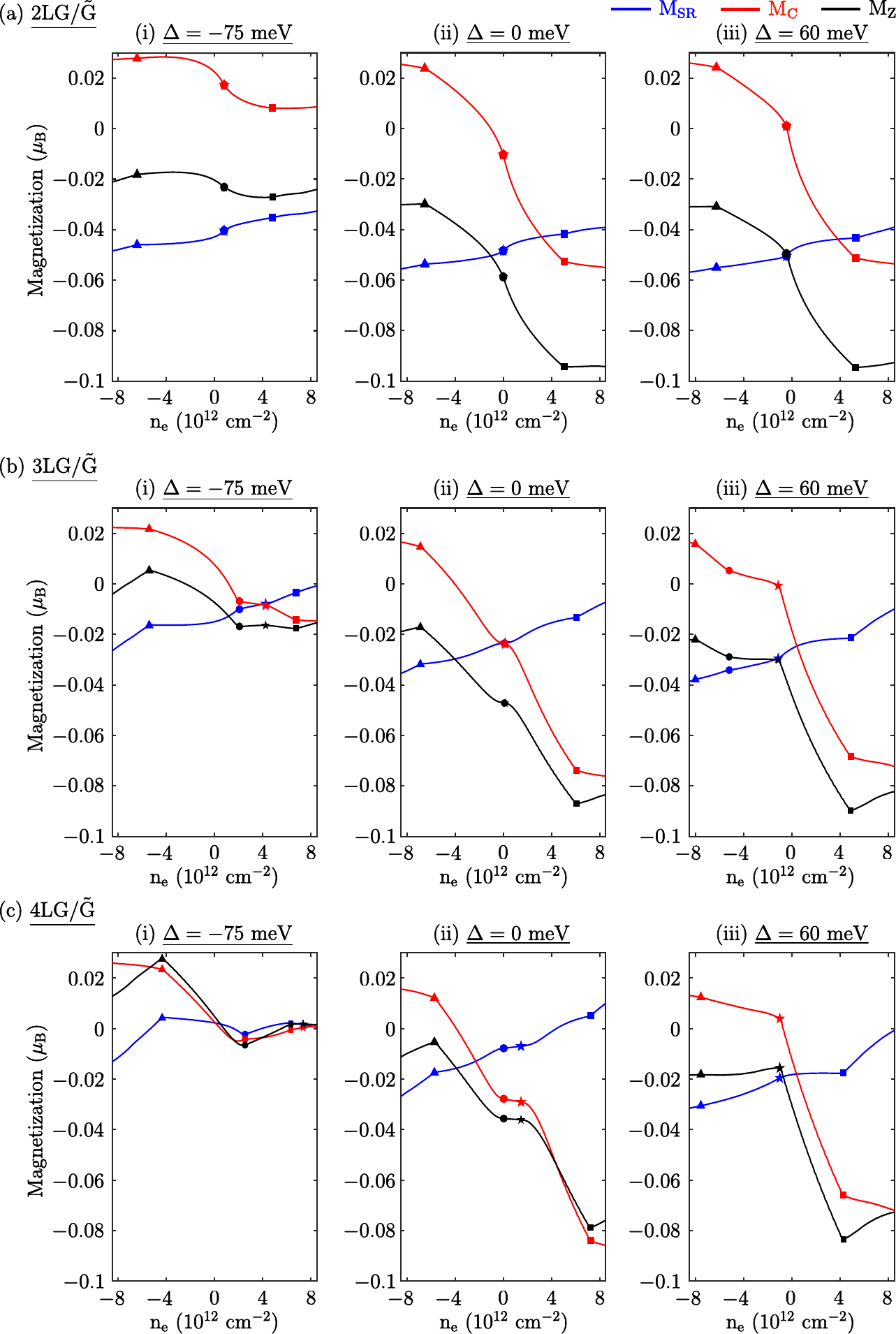}
    \caption{Decomposition of orbital magnetization in (a) 2LG/$\tilde{\mathrm{G}}$, (b) 3LG/$\tilde{\mathrm{G}}$, (c) 4LG/$\tilde{\mathrm{G}}$. The self-rotation ($M_{\mathrm{SR}}$, blue), center-of-mass ($M_C$, red), and total ($M_z^{\mathrm{orb}}$, black) moments versus carrier density for biases of (i) $\Delta=-75$~meV, (ii) $\Delta=0$~meV, and (iii) $\Delta=60$~meV.}
    \label{Decom_OrbMag}
\end{figure}

\begin{figure}[t]
    \centering
    \includegraphics[width=0.8\columnwidth]{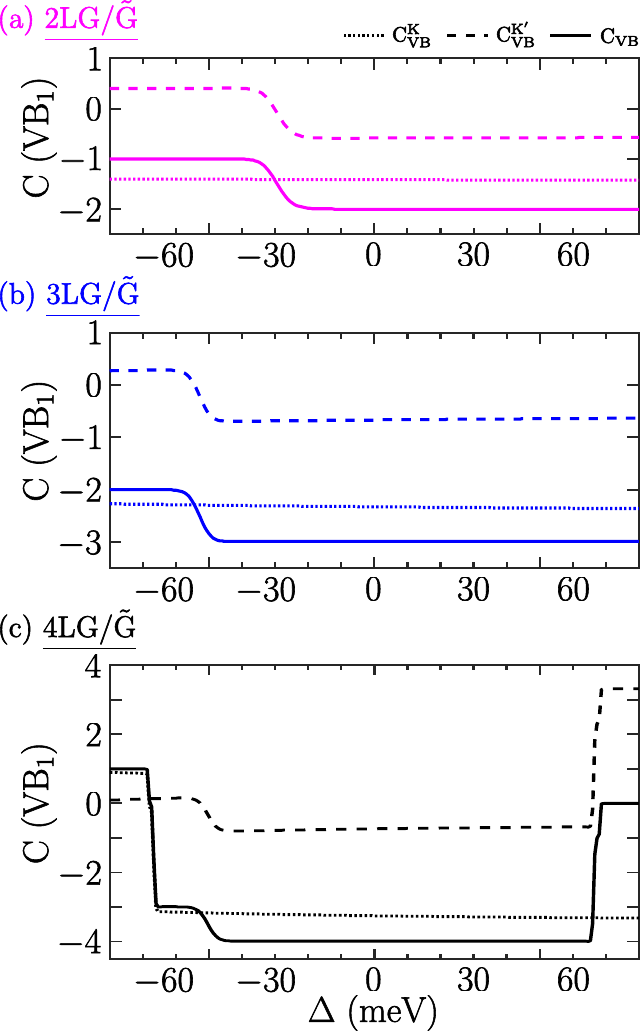}
    \caption{Chern number variation of the low-energy valence band ($\mathrm{VB_1}$) with interlayer potential $\Delta$ (meV) in (a) 2LG/$\tilde{\mathrm{G}}$, (b) 3LG/$\tilde{\mathrm{G}}$, (c) 4LG/$\tilde{\mathrm{G}}$.}
    \label{chern}
\end{figure}

The behavior of orbital magnetization becomes richer in multilayer graphene systems under Haldane proximity, due to the valley imbalance in the low-energy limit. This intrinsic feature gives rise to strong valley-dependent phenomena, with total magnetization $M_z^{\mathrm{orb}}(K,K^\prime) = M_z^{\mathrm{orb}}(K)+M_z^{\mathrm{orb}}(K^\prime)$~\cite{Ghosh_2025}. The interplay between $M_{\mathrm{SR}}$ and $M_C$ determines the total magnetization, and their relative contributions depend critically on the number of graphene layers, applied vertical electric field, and valley index. Since, time-reversal symmetry is broken, the system maintains an equivalent distribution of magnetic moments across valleys [Fig.~\ref{mz_line}(a)]. However, the magnitude of the magnetic moment near $K^\prime$ exceeds that near $K$, and this asymmetry is enhanced with increasing layer number and negative bias voltage [Fig.~\ref{mz_line}(b)]. This positive skew in the magnetic moment distribution drives the $M_{\mathrm{SR}}$ toward positive values in the low-density region [Fig.~\ref{Decom_OrbMag}(a)-(c)].
While both $M_{\mathrm{SR}}(K)$ and $M_{\mathrm{SR}}(K^\prime)$ remain negative regardless of doping, negative gate voltage significantly reduces $M_{\mathrm{SR}}(K^\prime)$ relative to $M_{\mathrm{SR}}(K)$ due to the positive moment distribution at the $K^\prime$ valley. Increasing the layer count to trilayer and tetralayer further suppresses the magnitude of $M_{\mathrm{SR}}$ noticeably [Fig.~\ref{Decom_OrbMag}(b,c)]. For $2\mathrm{LG}/\tilde{G}$ at zero interlayer potential, $M_{\mathrm{SR}}$ ranges from $-0.054$ to $-0.042\,\mu_B$ within the valley localized gap; this decreases to $-0.032$ to $-0.013\,\mu_B$ for trilayer and $-0.017$ to $0.005\,\mu_B$ for tetralayer, as highlighted in Fig.~\ref{Decom_OrbMag}(a-c). The sensitivity of $M_{\mathrm{SR}}$ to both doping and gate voltage is also enhanced in $3\mathrm{LG}/\tilde{G}$ and $4\mathrm{LG}/\tilde{G}$. Up to the $K^{\prime}$ valence band edge (triangle marker), total magnetization $M_z^{\mathrm{orb}}$ is dominated by $M_{\mathrm{SR}}(K^{\prime})$ [Fig.~\ref{Decom_OrbMag}]. In contrast, $M_C$ exhibits the opposite behavior. Due to the non-trivial topological valence band with negative Chern number, $M_C$ trends negative within the valley localized gap. Since the integrated Berry curvature of the $K$ valley ($C_{\mathrm{VB}}^K$) exceeds that of the $K^\prime$ valley ($C_{\mathrm{VB}}^{K^\prime}$) [Fig.~\ref{chern}], $M_C(K)$ varies more strongly than $M_C(K^\prime)$ across the low-density region. Near the $K^\prime$ valley localized gap, $M_C(K^\prime) > M_C(K)$, driving the total $M_C = M_C(K) + M_C(K^\prime)$ to small positive values ($\approx 0.025\,\mu_B$). As doping increases, $M_C(K)$ becomes dominant, pulling $M_z^{\mathrm{orb}}$ toward negative values. With increasing layer number, the variation in $M_C$ is slightly enhanced due to changes in the integrated Berry curvatures. Notably, gate-voltage-induced topological transitions modify the behavior of $M_C$. After the transition, $C_{\mathrm{VB}}^{K^\prime}$ becomes positive (Fig.~\ref{chern}), reversing the slope of $M_C(K^\prime)$ toward positive values. This partially compensates the dominance of $M_C(K)$, resulting in less volatile behavior within the valley localized gap, as shown in Fig.~\ref{Decom_OrbMag}(a-c.i).
\subsection{Magnetization Reversal}
\begin{figure*}[t]
    \centering
    \includegraphics[width=0.8\textwidth]{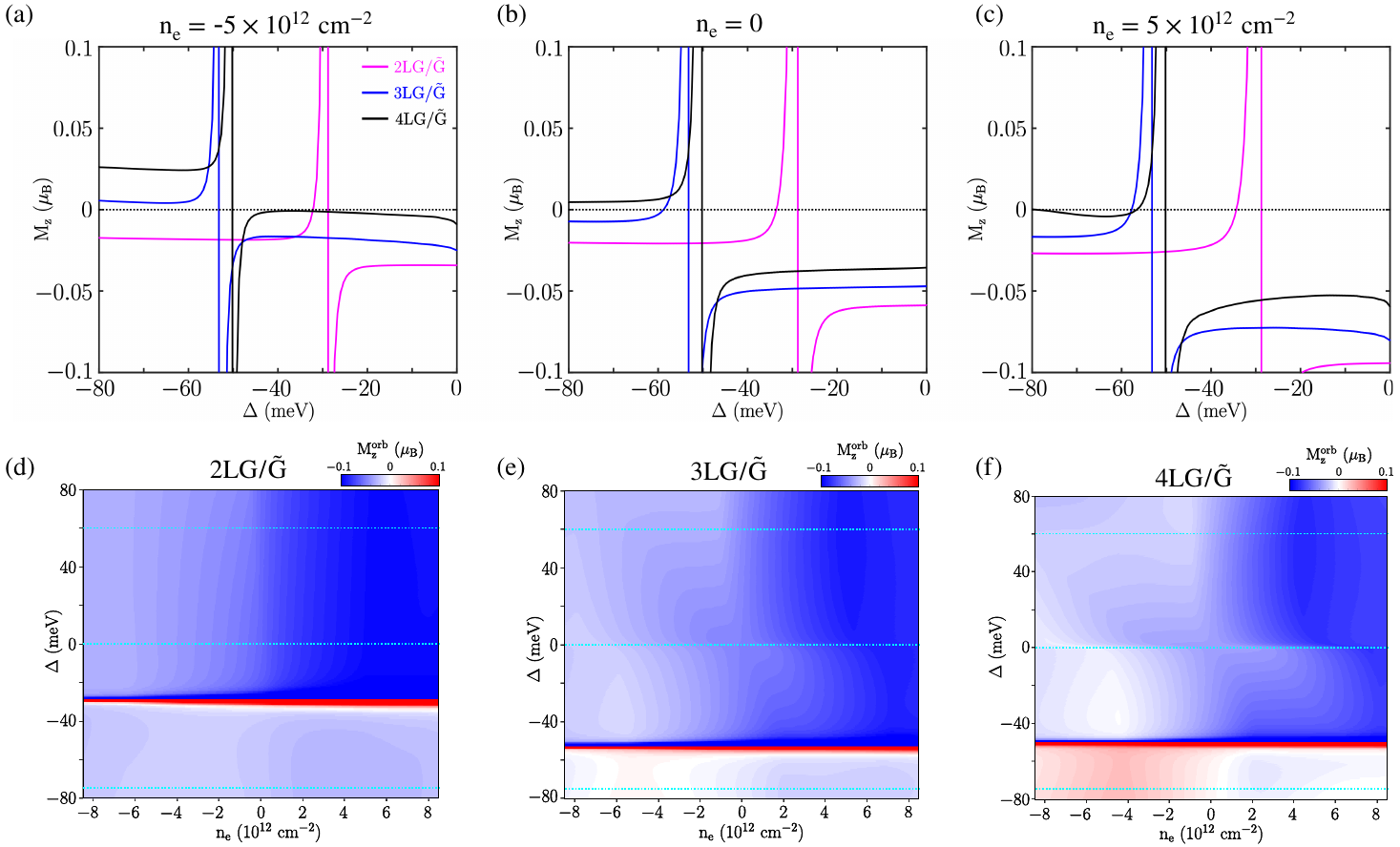}
    \caption{Orbital magnetization $M_z$ as a function of interlayer bias $\Delta$ for (a) hole-doped ($n_e = -5\times10^{12}\,\mathrm{cm^{-2}}$), (b) charge neutrality ($n_e = 0$), and (c) electron-doped ($n_e = +5\times10^{12}\,\mathrm{cm^{-2}}$) cases in bilayer (2LG/$\tilde{\mathrm{G}}$, magenta), trilayer (3LG/$\tilde{\mathrm{G}}$, blue), and tetralayer (4LG/$\tilde{\mathrm{G}}$, black) graphene under Haldane proximity. Orbital magnetization $M_z^{\mathrm{orb}}$ (color scale) as a function of interlayer bias $\Delta$ and carrier density $n_e$ for (d) bilayer, (e) trilayer, and (f) tetralayer graphene with Haldane proximity. Sky blue dashed lines indicate gate voltages of $-75$, $0$, and $60$ meV. Kinks in the magnetization, signaling topological phase transitions, are clearly visible. A distinct sign reversal of magnetization accompanies the transition in 3LG and 4LG.}
   \label{mag_density}
\end{figure*}
The orbital magnetization exhibits richer behavior in multilayer graphene under Haldane proximity and applied interlayer bias $\Delta$. Figure~\ref{mag_density} shows $M_z^{\mathrm{orb}}$ as a function of $\Delta$ and carrier density $n_e$ for bilayer, trilayer, and tetralayer graphene. A common feature across all systems is an abrupt change in $M_z^{\mathrm{orb}}$ coinciding with the closing and reopening of a valley-resolved band gap, signaling a topological phase transition. Despite this similarity, the magnetic response exhibits strong layer dependence. In bilayer graphene, $M_z^{\mathrm{orb}}$ remains negative and nearly monotonic, with only weak sensitivity to doping [Fig.~\ref{mag_density}(a-c)]. In contrast, trilayer and tetralayer graphene display a pronounced magnetization reversal at negative interlayer bias in the hole-doped regime ($n_e<0$). This doping-induced magnetic polarization flip occurs beyond critical biases of $\Delta\simeq -55$~meV for 3LG and $\Delta\simeq -50$~meV for 4LG, and is absent in the bilayer case [Fig.~\ref{mag_density}(d-f)]. This behavior originates from competition between $M_{\mathrm{SR}}$ and $M_C$. In the hole-doped regime near the $K^{\prime}$-valley valence band edge, $M_{\mathrm{SR}}$ dominates at low bias, yielding negative magnetization. As negative bias increases and layer number grows, $M_{\mathrm{SR}}$ is progressively suppressed while $M_C$ is enhanced. Beyond the critical bias, $M_C$ overcomes $M_{\mathrm{SR}}$, leading to magnetization reversal [Fig.~\ref{mag_density}(a)]. In 2LG/$\tilde{\rm G}$, although a kink appears at $\Delta \approx -30$~meV, the larger magnitude of $M_{\mathrm{SR}}$ prevents a sign change. In the electron-doped regime ($n_e > 0$), a strong enhancement of $M_z^{\mathrm{orb}}$ is observed without sign change, as $M_C$ increases linearly with chemical potential toward negative values. Throughout the doping region, $M_{\mathrm{SR}}$ and $M_C$ behave oppositely. $M_{\mathrm{SR}}$ remains constant within the valley-projected gap, while $M_C$ varies linearly with a slope determined by the occupied band Chern numbers. Although a global band gap is absent in Haldane-proximitized multilayer graphene, a valley-localized gap persists. When the Fermi level lies within this local gap of one valley while intersecting metallic bands of the opposite valley, the simple quantization rule is broken, enabling continuous and layer-dependent tuning of orbital magnetization. The presence of additional occupied metallic states allows $M_C$ to exceed $M_{\mathrm{SR}}$ within the valley localized gap. After the topological transition, magnetization reversal is directly tuned by $M_C$: positive $M_z^{\mathrm{orb}}$ induced by $M_C(K^\prime)$ changes polarity via $M_C(K)$ [Fig.~\ref{mag_density}(e-f)]. Importantly, the magnetization reversal is robust across both hole and electron doping, a unique finding absent in bilayer graphene [Fig.~\ref{mag_density}(d)]. Our predicted magnetization reversal should be observable through multiple complementary experimental techniques. In transport, at hole doping $n_e = -5 \times 10^{12}$ cm$^{-2}$, sweeping the displacement field through $\Delta \approx -55$ meV should produce a sign change in the anomalous Hall resistance, with hysteretic behavior indicating first-order switching between magnetic ground states~\cite{EPFL,Polshyn2020}. Thermodynamically, the sign reversal should manifest as a discontinuity in magnetic susceptibility measurable via torque magnetometry~\cite{BookFranco2021}. Spectroscopically, the accompanying change in valley polarization should be detectable as a shift in circular dichroism of optical absorption near the band edge~\cite{Cao2012}. These predictions identify critical displacement fields and doping levels where competing magnetic ground states should be most pronounced in trilayer graphene. Further, these results definitively establish layer number as a critical parameter for controlling the orbital magnetic response, with trilayer and tetralayer graphene uniquely enabling bias-driven magnetization reversal.

\subsection{Magnetic Response in $5LG/\tilde{G}$}
\begin{figure*}[t]
    \centering
    \includegraphics[width=0.85\textwidth]{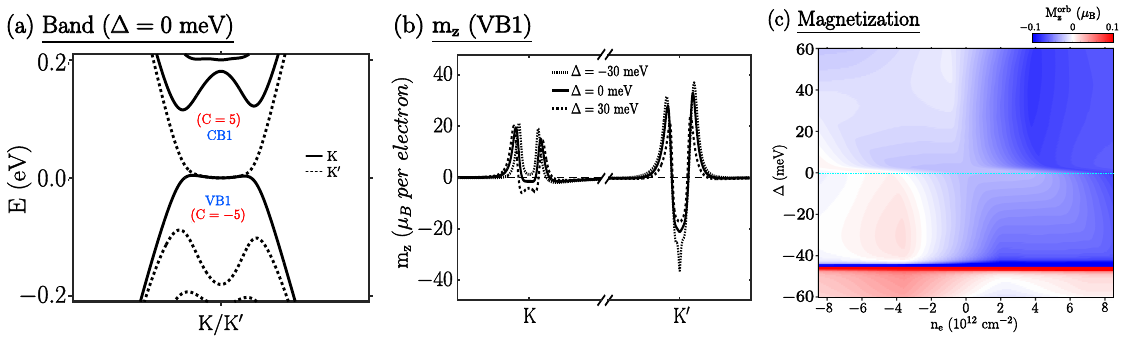}
    \caption{Pentalayer graphene on Haldane substrate. (a) Low energy bands at zero gate voltage. (b) Orbital magnetic moment distribution of VB1 at valleys K and $K^{\prime}$ and (c) total orbital magnetization as a function of density ($n_e$) and interlayer bias $\Delta$ }
    \label{5LG_H}
\end{figure*}
As discussed, the number of layers plays a key role in controlling orbital magnetization in graphene--Haldane heterostructures. In this section, we extend our study to multilayer rhombohedral graphene beyond four layers, focusing on pentalayer graphene proximitized by a Haldane substrate ($5\mathrm{LG}/\tilde{G}$). The low-energy bands (Fig.~\ref{5LG_H}(a)) show valley-dependent metallic behavior similar to the few-layer cases. Compared to $4\mathrm{LG}/\tilde{G}$, the magnetic moment $m_z$ at the $K$ valley in $5\mathrm{LG}/\tilde{G}$ is predominantly positive (Fig.~\ref{5LG_H}(b)). This substantial positive $m_z$ distribution near the valleys suppresses negative magnetic polarization even without an applied gate voltage. Figure~\ref{5LG_H}(c) shows how the total orbital magnetization varies with carrier density and gate voltage. Unlike in $2\mathrm{LG}/\tilde{G}$ or $3\mathrm{LG}/\tilde{G}$ or $4\mathrm{LG}/\tilde{G}$, magnetic reversal occurs even before the topological transition, requiring only a small hole density ($n_e < 0$) and a small negative gate voltage. In this low-hole regime, a small positive $M_C$ pushes the total orbital magnetization toward positive values. After the topological transition at $\Delta = -45$ meV, positive magnetic polarization dominates due to the combined effect of $M_{SR}$ and $M_C$. Upon further electron doping, however, the negative slope of $M_C$ brings back a negative total orbital magnetization.
This behavior, magnetic reversal occurring independently of the topological transition, mirrors the sensitivity of correlated topological phases recently observed in intrinsic RMG systems, where small changes in carrier density or electric field can stabilize fractional quantum anomalous Hall states or chiral superconductivity~\cite{Lu2024, Choi2025, Han2025, gm64-vxdm}. Our findings thus extend the notion of electrically tunable topology to proximitized RMG, offering a complementary platform that does not rely on moir\'e patterning.

\section{Conclusion}
We have systematically investigated the orbital magnetization in rhombohedral multilayer graphene (bilayer, trilayer, tetralayer and pentalayer) under Haldane proximity, using a tight-binding framework and the modern theory of orbital magnetization. Unlike monolayer graphene, which develops a global topological gap under Haldane proximity, multilayer systems remain metallic due to protected low-energy bands originating from unperturbed sublattices. Despite the absence of a global gap, these bands acquire non-trivial topology with layer-dependent Chern numbers. The orbital magnetic moment exhibits strong valley asymmetry with larger magnitudes near the $K^\prime$ valley compared to $K$. This asymmetry is enhanced by increasing layer number and negative interlayer bias, driving distinct contributions to the integrated magnetization. Decomposing the total orbital magnetization into self-rotation ($M_{\mathrm{SR}}$) and center-of-mass ($M_C$) contributions reveals their competing roles: $M_{\mathrm{SR}}$ remains relatively constant within valley-projected gaps but is progressively suppressed with increasing layer number and negative bias, while $M_C$ varies linearly with chemical potential and is enhanced in multi-layers.

\par Our most striking finding is a bias-induced sign reversal of orbital magnetization in trilayer and tetralayer graphene, occurring beyond critical negative biases in the hole-doped regime ($\Delta \simeq -55$ meV for 3LG, $-50$ meV for 4LG). This reversal is completely absent in bilayer graphene, where magnetization remains negative throughout. The effect originates from the crossover where $M_C$ overcomes $M_{\mathrm{SR}}$, enabled by the modified valley-specific Berry curvature distribution in multi-layer stacks. It is further established in the case of 5LG, that the magnetization reversal is independent of the topological transition, and depends on the direction of bias and hole doping. This is due to the low energy flat bands with the suppressed $M_{SR}$. Interestingly, in 3LG and 4LG graphene-Haldane heterostructures, the topological transition lead the $M_{\mathrm{C}}^{K^\prime}$ slope change and thus accelerated the magnetization reversal. While most pronounced in hole doping, the magnetization reversal also manifests in electron-doped regimes, indicating a robust mechanism rooted in the topological band structure rather than specific filling factors. Our results establish layer number as a critical control parameter for orbital magnetism in proximitized graphene systems. The ability to reverse orbital magnetization via electric field, without requiring moiré superlattices or magnetic fields, positions rhombohedral multilayer graphene as a promising platform for orbitronic and valleytronic applications. The magnetization magnitudes we obtain ($\sim 0.1\,\mu_B$ per electron) are comparable to those reported in twisted bilayer graphene~\cite{PhysRevB.102.121406} and other proximitized systems, placing our predictions within experimentally accessible regimes.

Several promising directions emerge from this work. Extending the analysis to pentalayer and hexalayer systems could reveal oscillatory layer-number dependence or higher-order topological phases. Incorporating electron-electron interactions may uncover competing correlated insulating states that hybridize with topological bands, enabling interaction-tuned magnetization reversal. Experimentally, our predictions are testable in graphene encapsulated between a transition metal dichalcogenide (providing proximity spin-orbit coupling) and a ferromagnetic insulator such as Cr$_2$Ge$_2$Te$_6$ (providing exchange field), with dual-gate control enabling independent tuning of doping and displacement field~\cite{ABC_CGT, WS2_AB_CGT, AB_WS2_CGT, AB-WSe2,science.adk9749}. Specifically, at hole doping $n_e = -5 \times 10^{12}$ cm$^{-2}$, sweeping the displacement field through $\Delta \approx -55$ meV should produce a sign change in anomalous Hall resistance, with hysteretic behavior indicating first-order switching between magnetic ground states. These predictions identify critical displacement fields and doping levels where competing magnetic orders should be most pronounced in trilayer graphene.
\section*{Acknowledgments}
We acknowledge the support provided by the Kepler Computing Facility, maintained by the Department of Physical Sciences, IISER Kolkata, for various computational requirements. S.G. acknowledges the support from the Council of Scientific and Industrial Research (CSIR), India, for the doctoral fellowship. B. L. C acknowledges the SERB with Grant No. SRG/2022/001102 and ``IISER Kolkata Start-up-Grant" No. IISER-K/DoRD/SUG/BC/2021-22/376.

\appendix
\section{Rhombohedral N-layer Graphene (NLG) Hamiltonian in momentum space}\label{app.NLG}
The single-particle tight-binding Hamiltonian in momentum space for the rhombohedral multilayer graphene in the $\{A_i,B_i\}$ sublattice basis is defined as~\cite{NLG_para_ref,Koshino2009} 
\begin{equation}
H^{NLG}(\vb{k})=
\begin{pmatrix}
    H_{11} & V & W & 0 & \cdots \\
    V^{\dag} & H_{22} & V & W & \ddots \\
    W^{\dag} & V^{\dag} & H_{33} & \ddots & \ddots \\
    0 & W^{\dag} & \ddots & H_{44} & \ddots  \\
    \vdots & \ddots & \ddots & \ddots
\end{pmatrix}_{N\times N}
\end{equation}
where the intralayer $(H_{ll})_{2\times2}(l=1,2,3,...,N)$ and interlayer $V_{2\times2}$ for the nearest layer and $W_{2\times2}$ for the next-nearest layer interaction terms are given by
\begin{equation}\nonumber
    H_{ll}(\vb{k})=
    \begin{pmatrix}
        u_{A_l} & t_0 f(\vb{k}) \\
        t_0 f^{\dag}(\vb{k}) & u_{B_l}
    \end{pmatrix}+ D_{ll} \mathbb{I}
\end{equation}
\begin{equation}
    V=\begin{pmatrix}
        t_4 f(\vb{k}) & t_3 f^{\dag}(\vb{k}) \\ t_1 & t_4 f(\vb{k})
    \end{pmatrix}
\end{equation}
\begin{equation}\nonumber
    W=\begin{pmatrix}
        0 & t_2 \\ 0 & 0
    \end{pmatrix}
\end{equation}
 The $D_{ll}\mathbb{I}$ term includes the inter-layer potential difference due to a applied vertical electric field through $D(ll)=(N-1)\frac{\Delta}{2}-(l-1)\Delta$, where we consider equal-magnitude potential drops ($\Delta$) across the consecutive NLG layers. The hopping parameters, namely, $(t_0,t_1,t_2,t_3,t_4)$ are used as (-3.1,0.3561,-0.0083,0.293,0.144) eV, similar to the literature's \cite{Ghosh_2025,NLG_para_ref} that are corresponding to the nearest neighbour intra-layer hopping terms between $A_l$ and $B_l$ and inter-layer $B_l$ and $A_{l+1}$, $A_l$ and $B_{l+2}$, $A_l$ and $B_{l+1}$ and $A_l(B_l)$ and $A_{l+1}(B_{l+1})$ sites respectively. The diagonal-site potentials $u_{A_l}(u_{B_l})$ at each sublattice are $u_{A_1}=u_{B_N}=0$, $u_{B_1}=u_{A_{N-1}}=0.0122$ eV and $u_{A_l/B_l}=-0.0164$ eV $(2<l<(N-1))$.
\bibliography{simple}
\end{document}